\documentclass[aps,prb, twocolumn]{revtex4-1}
\usepackage{filecontents}
\usepackage{amsmath}
\usepackage{amsfonts}
\usepackage{amssymb}
\usepackage{dcolumn}
\usepackage{bm}
\usepackage{graphicx}
\usepackage{adjustbox}
\usepackage{subcaption}
\usepackage{url}
\usepackage{xr-hyper}
\usepackage[hyperindex,breaklinks]{hyperref}
\usepackage{epstopdf}
\usepackage{color}
\usepackage{xcolor}
\usepackage{textcomp}
\usepackage{gensymb}
\usepackage{adjustbox}
\usepackage{ulem}
\usepackage{comment}
\usepackage{float}
\makeatletter
\newcommand*{\addFileDependency}[1]{
  \typeout{(#1)}
  \@addtofilelist{#1}
  \IfFileExists{#1}{}{\typeout{No file #1.}}
}
\makeatother

\newcommand*{\myexternaldocument}[1]{
    \externaldocument{#1}
    \addFileDependency{#1.tex}
    \addFileDependency{#1.aux}
}
\myexternaldocument{supplement}

\begin{document}
\title{A Combined First Principles Study of the Structural, Magnetic, and Phonon Properties of Monolayer CrI$_{3}$}
\author{Daniel Staros,$^{1, \dag}$ Guoxiang Hu,$^{2, 3, \dag}$ Juha Tiihonen,$^{6}$ Ravindra Nanguneri,$^{1}$ Jaron Krogel,$^{6}$ M. Chandler Bennett,$^{6}$ Olle Heinonen,$^{4,5}$ Panchapakesan Ganesh,$^{3, \dag\dag}$ and Brenda Rubenstein$^{1,\dag\dag}$}

\address{$^{1}$Department of Chemistry, Brown University, Providence, RI 02912, USA}
\address{$^{2}$Department of Chemistry and Biochemistry, Queens College, City University of New York, Flushing, NY 11367, USA}
\address{$^{3}$Center for Nanophase Materials Sciences Division, Oak Ridge National Laboratory, Oak Ridge, TN 37831, USA}
\address{$^{4}$Materials Science Division, Argonne National Laboratory, Argonne, IL 60439, USA}
\address{$^{5}$Northwestern-Argonne Institute for Science and Engineering, Northwestern University, Evanston, IL 60208, USA}
\address{$^{6}$Material Science and Technology Division, Oak Ridge National Laboratory, Oak Ridge, TN 37831, USA}

\date{\today}

\begin{abstract}
The first magnetic 2D material discovered, monolayer (ML) CrI$_3$, is particularly fascinating due to its ground state ferromagnetism. Yet, because monolayer materials are difficult to probe experimentally, much remains unresolved about ML CrI$_{3}$'s structural, electronic, and magnetic properties. Here, we leverage Density Functional Theory (DFT) and high-accuracy Diffusion Monte Carlo (DMC) simulations to predict lattice parameters, magnetic moments, and spin-phonon and spin-lattice coupling of ML CrI$_{3}$. We exploit a recently developed surrogate Hessian DMC line search technique to determine CrI$_{3}$'s monolayer geometry with DMC accuracy, yielding lattice parameters in good agreement with recently-published STM measurements-- an accomplishment given the $\sim10$\% variability in previous DFT-derived estimates depending upon the functional. Strikingly, we find previous DFT predictions of ML CrI$_3$'s magnetic spin moments are correct on average across a unit cell, but miss critical local spatial fluctuations in the spin density revealed by more accurate DMC. DMC predicts magnetic moments in ML CrI$_3$ are 3.62 $\mu_B$ per chromium and -0.145 $\mu_B$ per iodine; both larger than previous DFT predictions. The large disparate moments together with the large spin-orbit coupling of CrI$_3$'s I-$\sl p$ orbital suggests a ligand superexchange-dominated magnetic anisotropy in ML CrI$_3$, corroborating recent observations of magnons in its 2D limit. We also find ML CrI$_3$ exhibits a substantial spin-phonon coupling of $\sim$~3.32 cm$^{-1}$. Our work thus establishes many of ML CrI$_{3}$'s key properties, while also continuing to demonstrate the pivotal role DMC can assume in the study of magnetic and other 2D materials. 
\end{abstract}

\maketitle

\section{Introduction \label{sec-introduction}}
  \vspace*{-0.20cm}
 
2D materials represent an exciting new frontier for materials research.\cite{Novoselov_Science} Due to their reduced dimensionality, these materials tend to exhibit stronger and longer-range electron correlation than their 3D counterparts that gives rise to exotic new physics and phase behavior, including Moiré patterns,\cite{Tang_AdvFuncMats, He_ACSNano} two-dimensional superconductivity,\cite{Qiu_AdvMats} and exotic spin and charge density waves.\cite{Chen_NatureComm, Huang_PRL, Wang_ACSNano} The properties of 2D materials can also be tuned just by stacking,\cite{Guo_AdvFunctMater, Sivadas_NanoLett, Soriano_SSComm} crinkling,\cite{Chen_NanoscaleHoriz} straining\cite{Dai_AdvMater, Song_NatureMater, Li_NatureMater}, and twisting\cite{trambly2010,bistritzer2011,dos2012,yankowitz2019} them. This versatility facilitates the layer-by-layer construction of designer materials with unique properties through the careful selection and ordering of their constituent layers.\cite{Zeng_ChemReviews} 

An exciting recent development in this regard is the discovery of new magnetic 2D materials.\cite{Gibertini_NatureNano} While the Mermin-Wagner Theorem,\cite{Mermin_PRL, Hohenberg_PhysRev} prohibits finite-temperature magnetism for the isotropic Heisenberg model in 2D, magnetic 2D materials such as  CrI$_3$,\cite{Huang_NatureLett}
Fe$_3$GeTe$_2$,\cite{fei2018} and VSe$_2$\cite{Wang_ACSNano2} have large anisotropies arising from strong spin-orbit couplings or structural anisotropies that break continuous symmetries and allow for finite-temperature long-range magnetic order. 2D magnetic materials are of great interest as they may accelerate the development of next-generation spintronic data transmission\cite{Behera_AppPhysLett} and storage\cite{Liu_AIP} technologies. Furthermore, if integrated into 2D heterostructures, magnetic 2D materials may additionally be able to induce magnetism in nearby nonmagnetic layers through proximity effects, enabling the realization of magnetic graphene,\cite{Karpiak_2DMater}.  Few-layer magnetic materials  that have particularly strong magneto-optical responses and/or electron-phonon coupling, e.g., \textit{M}PSe$_3$/TMD heterointerfaces,\cite{Onga_Nano} with $M$=Mn, Fe , WSe$_2$/CrI$_3$ heterostructures,\cite{Mukherjee_NatureComm} or bilayer CrI$_3$\cite{Jin_NatureComm} are also potentially promising for applications that bridge spintronics, photonics, and phononics.

However, in order to realize the potentials of 2D materials, their structural, electronic, magnetic, and optical properties have to be well characterized and understood. Experimentally, this is very challenging for mono-and few-layer systems. Monolayers are often grown on substrates that can distort their intrinsic geometries.\cite{Li_ScienceBull, Ahn_NatComm, Yan_PCCP} Due to their inherently small thickness, monolayers are also not immediately accessible to neutron and other scattering experiments, which typically require a critical thickness to yield meaningful scattering patterns.\cite{Velicky_apmt} Given these experimental challenges, it is imperative to predictively and accurately model these materials' atomic geometries, spin densities, and magnetic moments using first principles methods in order to elucidate the origins of their magnetic anisotropy and also to inform models upon which more macroscopic predictions can be built.\cite{Paul_NatComm, Meyer_NatComm} Recent advances in optical response-based methods,  such as fluorescence and magneto-optical Kerr effect (MOKE) microscopy, can resolve magnetic moments down to sub-micrometer scale,\cite{soldatov2018} and single-spin microscopy can improve this resolution down to nanometer scale,\cite{Thiel_Science} complementing such first principles-based predictions. 

To date, the overwhelming majority of first principles modeling of these materials has relied on Density Functional Theory (DFT).\cite{parr1989density} Even though DFT has led to transformative changes in our understanding of materials and chemistry and typically yields good results for ground state properties in 3D, especially lattice constants, it can yield widely varying results for the properties of 2D materials, such as lattice parameters and band gaps, depending upon the functional employed. Recent studies of 2D materials have shown that DFT consistently overestimates interlayer binding energies between 2D monolayers\cite{Wines_JCP_2020} and predicts lattice parameters that routinely differ from experimental ones by several percent.\cite{Mostaani_PRL, Shin_PhysRevMats} In contrast, Diffusion Monte Carlo (DMC), a real-space, many-body quantum Monte Carlo method, routinely predicts lattice constants within 1\% and band gaps to within 10\% of their experimental values for the same materials.\cite{Mostaani_PRL,Wines_JCP_2020,Shin_PhysRevMats,Shulenburger_NanoLett} It is worth emphasizing that DMC obtains such high-accuracy results for all properties within a single, consistent framework with few approximations; moreover, the approximations can be systematically improved upon because of the variational nature of DMC. In contrast, obtaining accurate DFT results of different properties often necessitates using a different functional for each and every property, which substantially reduces the predictive power of such calculations. 

\begin{figure}[t]
 \includegraphics[width=3.3 in]{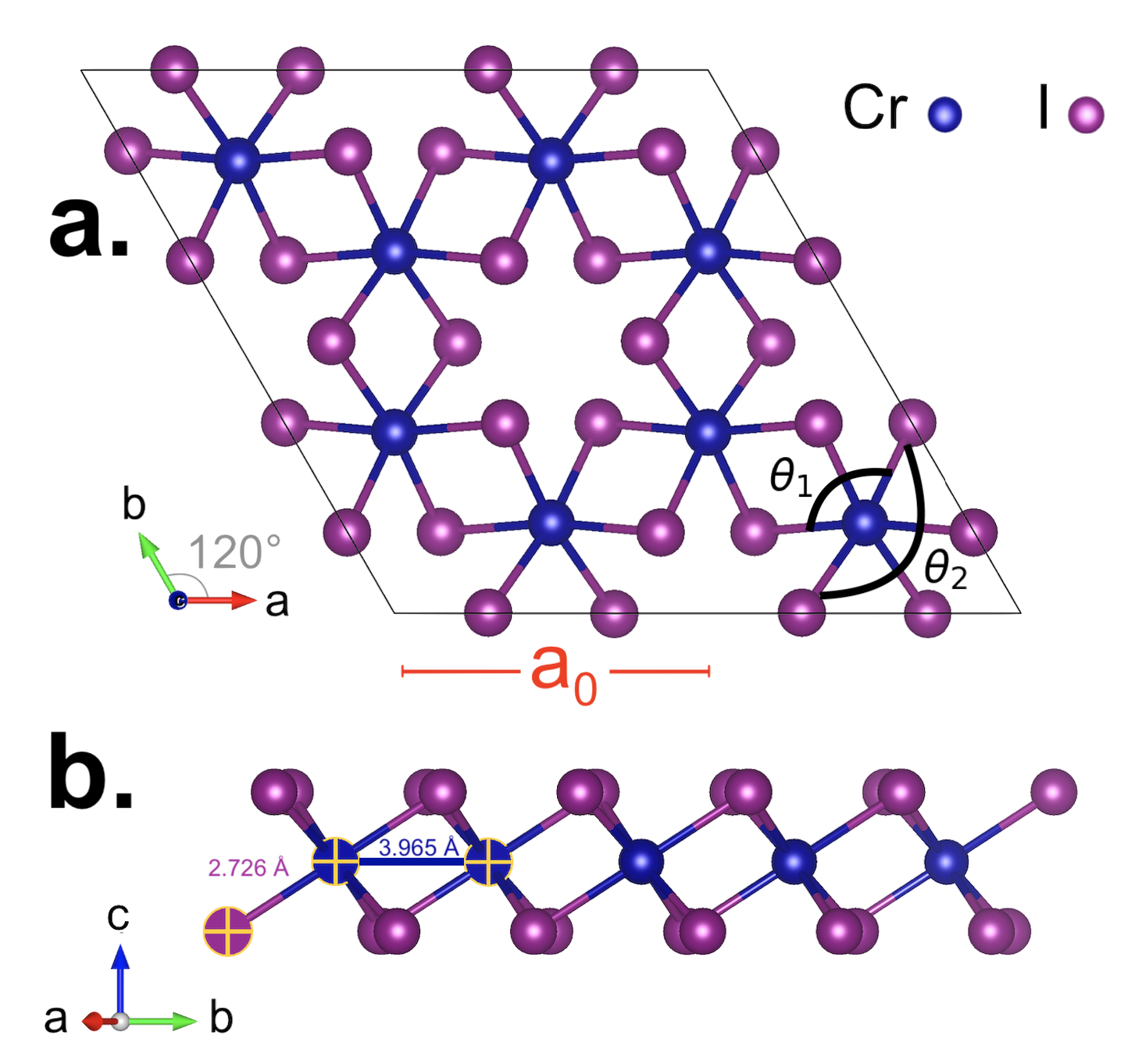}
 \caption{Geometry of monolayer CrI$_3$ cleaved from the bulk structure reported in reference:\cite{McGuire_ChemMater} (a) Top view depicting lattice constant of $a_0 = 6.867$ Å and the bond angles $\theta_1$, $\theta_2$ computed in this work. (b) Side view depicting the Cr$_1$-I bond distance of $2.726$ Å (purple) and Cr$_1$-Cr$_2$ bond distance of $3.965$ Å (blue).} 
 \label{fig:mock_structure}
\end{figure}

In this work, we use a combination of DFT and DMC\cite{Kim_2018} methods to construct a full, atomistic picture of the physics and properties of the CrI$_3$ monolayer (ML). ML CrI$_3$ is a strongly-correlated, magnetically-tunable Mott insulator with a hexagonal lattice structure in which each Cr atom is coordinated by six I atoms to form a distorted octahedron (see Figure \ref{fig:mock_structure}).\cite{McGuire_ChemMater} Atomic force microscopy and scanning tunneling microscopy point to a lattice constant of roughly 7~\r{A} \cite{Li_ScienceBull} and a monolayer thickness of 0.7~nm.\cite{Huang_NatureLett, Li_ScienceBull} What makes ML CrI$_3$ particularly intriguing as a material is its magnetism. In its monolayer form, CrI$_3$ has been demonstrated by MOKE microscopy\cite{Huang_NatureLett} and single-spin nitrogen vacancy microscopy\cite{Thiel_Science} experiments to be a ferromagnet. The strong magneto-crystalline anisotropy with an out-of-plane easy axis makes the Ising model a natural starting point for modeling the magnetic properties of CrI$_3$ as it allows for finite-temperature long-range order in 2D.\cite{Mermin_PRL} However, several DFT studies have suggested that the material may be better described by the Heisenberg and Kitaev models.\cite{Lado_PhysRevB,Xu_npj} These studies concluded that the ligand superexchange within slightly distorted Cr-I crystal environments\cite{Lado_PhysRevB} and spin-orbit coupling\cite{Xu_npj} ultimately stabilize ML CrI$_3$'s ferromagnetism. Even so, one DFT study of bilayer (BL) CrI$_3$ found non-negligible magnetic moments present on the iodine atoms,\cite{Besbes_PRB} suggesting that a more comprehensive magnetic model may be needed to quantify all of the relevant exchange interactions. 
In addition to its magnetism, CrI$_3$ exhibits strong electron-phonon coupling and magnetoelasticity, as evidenced by its alternating ferromagnetic odd-layered structures/antiferromagnetic even-layered structures\cite{Sivadas_NanoLett} and magnetic ordering-dependent Raman spectra.\cite{Zhang_NanoLett} Although computational studies of ML CrI$_3$'s properties have previously been performed using a variety of DFT functionals, including the generalized gradient approximation (GGA),\cite{McGuire_ChemMater,Webster_PhysRevB, Zhang_JMaterChemC}, hybrid functionals such as  Heyd-Scuseria-Ernzerhof  (HSE06),\cite{Zhang_JMaterChemC} and the Local Density Approximation (LDA) with added spin-orbit coupling (SOC) corrections (LDA$_{SOC}$),\cite{Webster_PhysChemChemPhys} they have yielded conflicting information regarding magnetic moments, lattice constant, and exchange parameters because of the strong functional dependence of these properties (see Supplementary Table \ref{tab:parameter_table2}). These discrepancies underscore the need to employ many-body approaches beyond DFT with fewer or no uncontrolled approximations in order to faithfully resolve ML CrI$_3$'s remaining controversies and uncertainties. 

We utilize a combination of DFT with an on-site Hubbard-U correction (DFT+$U$) as well as many-body DMC simulations to resolve the structural, magnetic, and phonon properties of monolayer CrI$_3$. Our results show that previous DFT predictions of ML CrI$_3$'s magnetic spin moments\cite{Besbes_PRB} are correct on average across a unit cell, but miss critical local spatial fluctuations and anisotropies in the spin density. Our DMC calculations predict a magnetic moment of 3.62 $\mu_B$ per chromium atom and -0.145 $\mu_B$ per iodine atom, which are considerably larger than past estimates in the literature. We furthermore exploit a recently developed surrogate Hessian DMC line search technique\cite{Shin_PhysRevMats} to determine CrI$_{3}$'s monolayer geometry with DMC accuracy, yielding high-accuracy bond lengths and bond angles that resolve previous structural ambiguities. Lastly, our calculations also reveal that ML CrI$_3$ possesses a substantial spin-phonon coupling, approximately  $3.32$~cm$^{-1}$, in line with couplings recently observed in other magnetic 2D materials. These findings indicate a far more anisotropic spin density involving more substantial ligand magnetism than previously thought.   

In Sec.~\ref{methods}, we describe the DFT and quantum Monte Carlo methods used, and in Sec.~\ref{Results} we present results on the monolayer's structure, spin density, magnetic moments, and phonon spectra. Finally, Sec.~\ref{Conc} contains a summary and conclusions.

\section{Computational Approach \label{methods}}

The ground state properties of the CrI$_3$ monolayer, including its lattice constant and magnetic moments, were modeled using DFT+$U$, Variational Monte Carlo (VMC), and DMC. DFT simulations were first performed to obtain reference structural data, guide our surrogate Hessian line search, and generate DFT+$U$ wave functions that were subsequently used as inputs into progressively more accurate VMC and DMC simulations. VMC and DMC calculations were then performed to obtain CrI$_3$'s  geometry, spin density, and magnetic moments.  In addition,  DFT+$U$ simulations with a self-consistent Hubbard-U\cite{cococcioni2005linear} were employed to investigate the effects of long-range magnetic ordering on lattice phonons.

\subsection{DFT Simulations \label{DFTUMeth}} 
DFT calculations were performed using the Vienna \textit{Ab Initio} Simulation Package.\cite{Kresse_PhysRevB, Kresse_PhysRevB2, Kresse_CompMatSci, Kresse_PhysRevB3} The ion-electron interaction was described with the projector augmented wave (PAW) method.\cite{Blochl_PhysRevB, Kresse_PhysRevB4} A cutoff energy of 500~eV was used for the plane-wave basis set. The Brillouin zone was sampled with an 8 × 8 × 1 Monkhorst-Pack {\bf k}-point mesh for the unit cell of CrI$_3$, which includes two Cr and six I atoms. A wide range of exchange correlation (XC) functionals were used to investigate the electronic and magnetic properties of the CrI$_3$ monolayer. The Local Density Approximation (LDA), Perdew–Burke–Ernzerhof (PBE), PBE+$U$, and PBEsol\cite{Perdew_PhysRevLett} functionals were specifically employed in this study. All calculations were spin-polarized. In its ferromagnetic (FM) configuration, all of CrI$_3$'s magnetic moments were initialized in the same out-of-plane direction, while in its antiferromagnetic (AFM) configuration, the two Cr atoms per unit cell were set to have antiparallel spins. A large vacuum of more than 25 \r{A} along the z-direction was employed to avoid artificial interactions between images. The energies were converged with a 1×10$^{-8}$ eV tolerance. In addition to screening various XC functionals, a self-consistent Hubbard $U$ ($U_{\rm LR}$) was determined from first principles by using the linear response approach proposed by Cococcioni and Gironcoli,\cite{cococcioni2005linear} in which $U$ was determined by the difference between the screened and bare second derivative of the energy with respect to localized state occupations. The linear-response calculation was also performed using Quantum Espresso, using the same high-quality, but hard,  pseudopotentials as were used to generate trial wave functions for the Quantum Monte Carlo simulations (discussed below). We obtained a linear-response $U$ value ($U_{\rm LR}$) value of $\sim$ 3.3 eV for Cr with the LDA functional. 

\subsection{Variational and Diffusion Monte Carlo Simulations \label{MCMeth}}

\begin{figure}[b]
 \includegraphics[width=3.3 in]{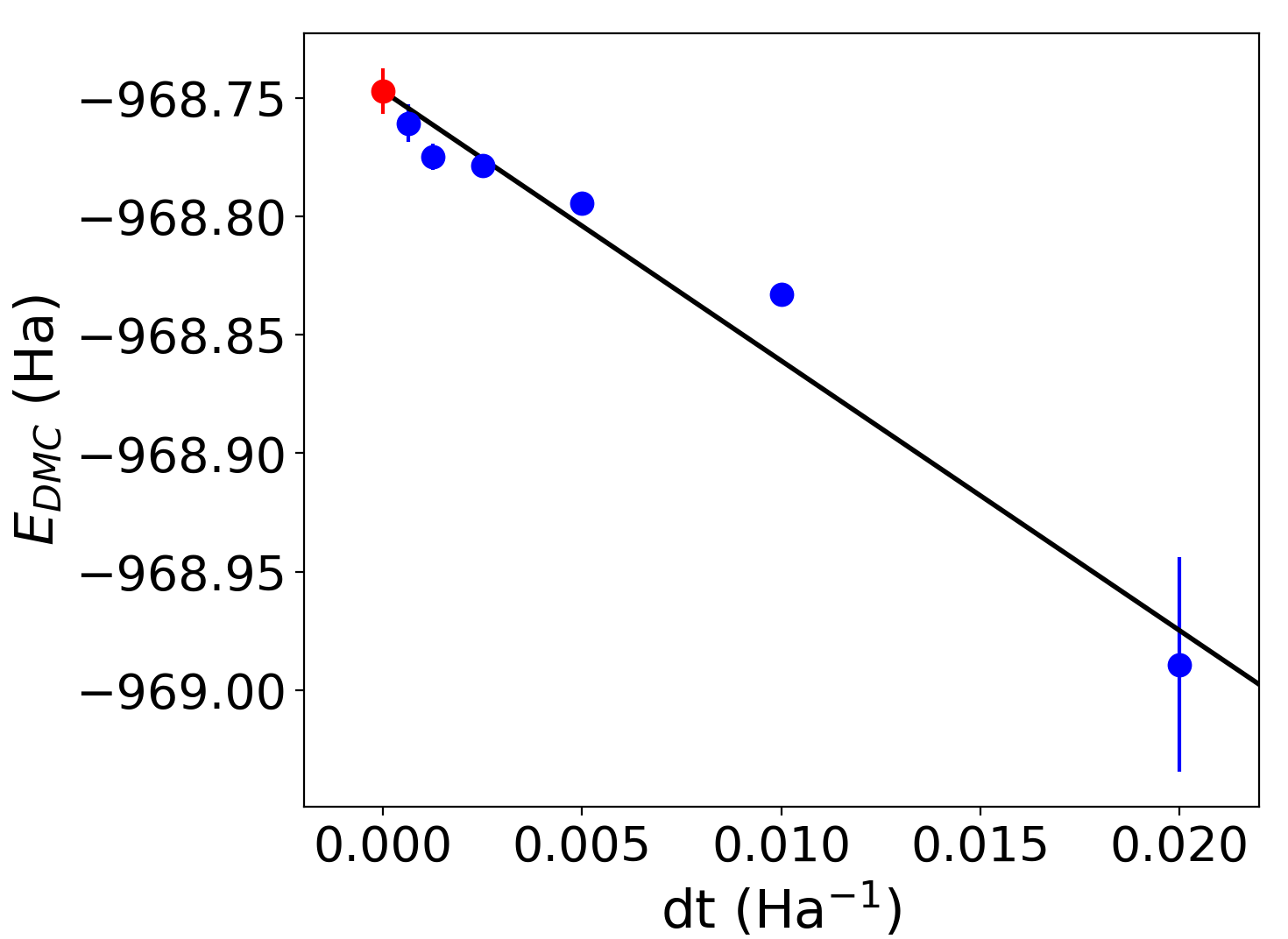}
 \caption{Timestep extrapolation of DMC energies obtained for a 2x2x1-tiled primitive cell of monolayer CrI$_3$. The blue points represent DMC energies and their accompanying error bars computed for the given time step sizes (dt), while the red point and its error bar denote the extrapolated 0-timestep energy based upon a linear fit of the data. The black line denotes the best linear fit to the data.} 
 \label{fig:Monolayer_timestep}
\end{figure}

VMC and DMC calculations were undertaken using QMCPACK,\cite{Kim_2018} a process that was facilitated by a Nexus workflow.\cite{Krogel_CompPhysComm} All QMC calculations were performed using a norm-conserving scalar-relativistic Opium-generated Cr pseudopotential and a norm-conserving non-relativistic iodine pseudopotential that was developed by Burkatzki, Fillipi and Dolg.\cite{Burkatzki_JChemPhys} Comparison of DMC energies obtained using nonrelativistic and relativistic iodine pseudopotentials yielded an energy difference of 0.009~meV/f.u., which was deemed negligible for this study. Both pseudopotentials were of the Troullier-Martins flavor. Trial wave functions used as starting points for the VMC calculations were produced using LDA+$U$ with $U=2$~eV within Quantum Espresso. In the DFT calculations, a plane-wave energy cutoff of 300~Ry was necessary to converge the total energy due to the high-quality, but hard pseudopotentials. The trial Jastrow factor consisted of inhomogeneous one- and homogeneous two-body terms, each represented by sums of 1D B-spline-based correlation functions in electron-ion or electron-electron pair distances. The trial Jastrow factor was optimized using the linear method. The VMC energy and variance were converged at a B-spline meshfactor of 1.2 using the fixed optimized Jastrow factor.

\begin{figure}[t]
 \includegraphics[width=3.3 in]{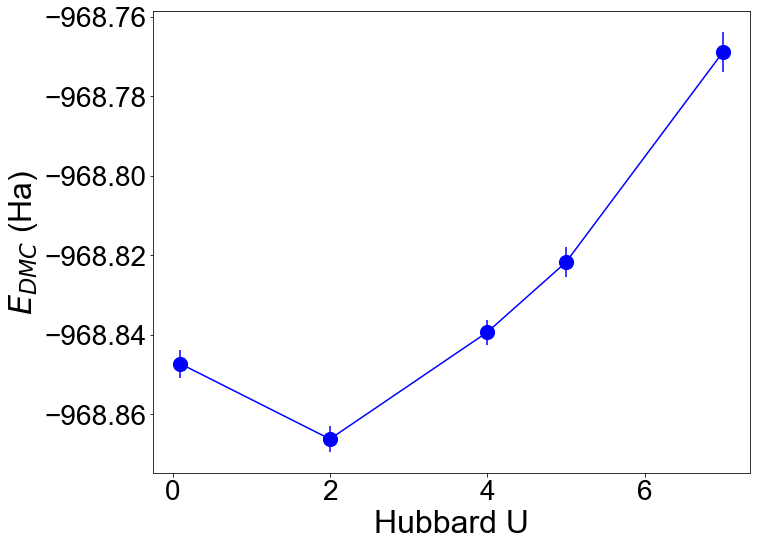}
 \caption{Twist-averaged DMC energies for a 2x2x1 supercell of monolayer CrI$_3$ obtained using LDA+$U$ trial wave functions with varying values of $U$. The minimum energy occurs around $U=2$ eV, suggesting that a LDA+$U$ trial wave function with $U=2$ would be the best to use in production-level DMC calculations. 
 }
 \label{fig:Monolayer_uValues}
\end{figure}

To determine a timestep that is a reasonable compromise between speed and accuracy, as well as the $U$ in LDA+$U$ trial wave functions that minimizes fixed-node errors, preliminary calculations were performed on a 2×2×1 supercell. These calculations demonstrated that a timestep of 0.01 Ha$^{-1}$ is capable of converging the DMC energies to within 0.01 Ha per formula unit (see Figure \ref{fig:Monolayer_timestep}). Moreover, as depicted in Figure \ref{fig:Monolayer_uValues}, the lowest DMC energies were achieved using a LDA+$U$ trial wave function with $U=2$ eV. A timestep of 0.01 Ha$^{-1}$ and LDA+$U$ trial wave functions with $U=2$ eV were thus used in all of our subsequent DMC production runs. Note that the obtained optimal value of $U$ from DMC is close to that of the $U_{LR} \sim 3.3$ eV value obtained from linear-response calculations. Further, both values agree with statistically indistinguishable total energies from a recent DMC study of bulk CrI$_3$ (see Fig. 1 in Ref. \onlinecite{Ichibha_PRM_2021}). All of our calculations also utilized a 3×3×1 QMC twist grid and a meshfactor of 1.2.

\begin{figure}[t]
 \includegraphics[width=3.45 in]{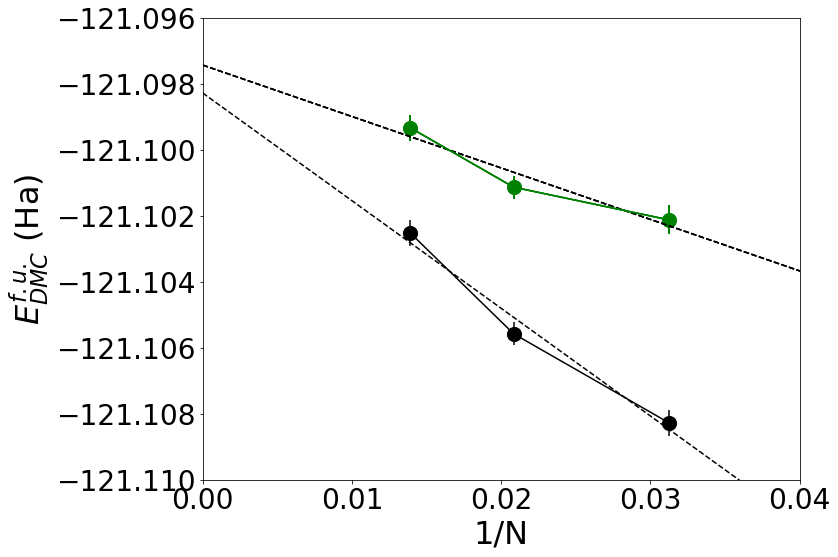}
 \caption{Finite size extrapolation of DMC energies per formula unit (f.u.) to their thermodynamic limit using 32-(2x2x1), 48-(2x3x1), and 72-(3x3x1) atom supercells. The green circles denote energies corrected using model periodic Coulomb (MPC) finite size corrections, while the black circles denote uncorrected energies. Note that the finite size-extrapolated energies obtained with and without corrections are within 2 mHa of one another. 
 }
 \label{fig:fs_extrapolation}
\end{figure}

VMC calculations employing twist-averaged boundary conditions were performed for increasingly larger twist grids and were shown to converge with a 3x3x1 twist grid; this size twist grid was thus used for twist-averaging of all subsequent VMC and DMC calculations to minimize one-body finite size effects.\cite{Ceperley_PRE} We further employed extrapolation to the thermodynamic limit of Model Periodic Coulomb-corrected and uncorrected DMC energies for increasingly larger supercells with 1×2×1 (16 atoms), 2×2×1 (32 atoms), 2×3×1 (48 atoms) and 3×3×1 (72 atoms) tilings to minimize the two-body finite-size errors (Figure \ref{fig:fs_extrapolation}).\cite{Holzmann_PRB, Kim_2018} All DMC runs used the T-moves scheme for pseudopotential evaluation to minimize localization errors.\cite{Casula_JChemPhys, Dzubak_JChemPhys}

\subsection{Calculation of Magnetic Moments via Diffusion Monte Carlo \label{MomentMeth}}
To predict the site-averaged atomic magnetic moment per chromium, $M_{Cr}$, and iodine, $M_{I}$, the spin densities, $\rho_s(r)$, obtained from both our DFT and DMC simulations were integrated up to a cutoff radius $r_{cut}$, defined as the zero-recrossing radius of the sign of the spin density (see Figure \ref{fig:rcut_figure}). In particular, we sum over the spherically-interpolated spin densities within a 16-atom unit cell to predict the magnetic moment per atom ($M_A$)
\begin{equation}
	M_{A} = 4 \pi \int_0^{r_{cut}} r^2 \; \rho_s (r) \; dr \approx 4\pi \sum_{i=0}^{r_{cut}/dr} r_i^2 \; \rho_s (r_i) dr,  
\end{equation}
where the $r_{i}$ denote the distances from the center of the atom to the given point on the grid. To facilitate a direct comparison between the DFT and DMC spin densities, the spin densities obtained from DFT simulations were interpolated onto the dimensions of the DMC spin density grid, before being mapped onto a spherical grid. 

\subsection{DMC Geometry Optimization through a Surrogate Hessian Line Search Method \label{Hessian}} 

In order to predict highly accurate lattice parameters for ML CrI$_3$, we used a Surrogate Hessian-Accelerated, DMC energy-based structural optimization method which has recently demonstrated robust performance when applied to two-dimensional GeSe.\cite{Shin_PhysRevMats} This method leverages the DFT energy Hessian, which is substantially cheaper to compute than the DMC energy Hessian, to direct where to compute DMC energies based on optimal statistical sampling to ultimately determine the lowest-energy material geometry. This line search method resolves structural parameters to an accuracy higher than is obtainable via DFT energy gradient-based structural optimization techniques; this resolution is particularly important for materials which exhibit sensitive coupling between electronic, magnetic, and structural degrees of freedom, such as 2D CrI$_3$. 

Our line search begins by first constructing an approximate energy Hessian ($H_p$), or force constant matrix, using a cheaper theory, which in our case is DFT. This step is meant to: 1) address the expense that would result from exploring an arbitrarily complex PES from scratch using only QMC energies and 2) minimize the noise that would result from numerous stochastic energy evaluations by allowing for a few, optimally placed QMC energy calculations. We expanded the DFT PES to second order in Wyckoff parameter space, $p$, as in Ref. \onlinecite{Krogel_prep}
\begin{equation}
    E(p) = E_0+\frac{1}{2}(p-p_0)^T H_p (p-p_0)
\end{equation}
and diagonalized the Hessian to obtain search directions conjugate to the PES isosurfaces 
\begin{equation}
    E(p) = E_0+\frac{1}{2}(p-p_0)^T U^T M U (p-p_0),
\end{equation}
where the columns of $U^T$ form an optimal basis of parameter directions for the line search.

With these search directions, a set of structures containing an equilibrium structure and a total of eighteen ``strained'' structures is generated which comprise the structure population for the first line search iteration. More specifically, three increasingly ``positively'' (tensile) strained structures and three increasingly ``negatively'' (compressive) strained structures are generated along each search direction by systematically varying the corresponding parameters of the equilibrium structure. Next, DMC energies are obtained for the entire structure population. The shape of the resultant DMC PES is then used to locate a candidate minimum energy structure.  The next iteration of the line search is performed identically to the first, but starting from this new candidate minimum structure. This process is repeated until the statistical noise on the parameter error bars are within the desired limit.\cite{Krogel_prep} Like all conjugate direction methods, a property of the method is the potential to converge after only a single iteration within a suitably quadratic region of the PES and using ideal search directions. In this work, converged DMC-based structural parameters were obtained after only three iterations, and were averaged over the last two iterations to account for remnant statistical fluctuations in the lattice constant.

\subsection{Phonon Calculations \label{phonons}} 
Phonon calculations meant to examine monolayer CrI$_3$'s spin-phonon coupling  were performed using the frozen phonon method as implemented in the PHONOPY code\cite{phonopy} based upon DFT calculations performed in VASP.\cite{Kresse_PhysRevB3,Kresse_PhysRevB4} We explored how CrI$_{3}$'s phonons change with different magnetic orderings, including nonmagnetic, ferromagnetic, and antiferromagnetic orderings, using both the LDA and LDA+$U$=3 eV functionals. Our choice of $U=3$ eV is further motivated below. 2x2x1 supercell structures with displacements were created from a unit cell fully preserving the material's crystal symmetry. Force constants were calculated using the structure files from the computed forces on the atoms. A part of the dynamical matrix was built from the force constants. Phonon frequencies and eigenvectors were calculated from the dynamical matrices with the specified {\bf q}-points. Monolayer CrI$_3$ possesses D$_3d$ point group symmetry and hence the phonon modes at the $\Gamma$ point can be decomposed as G$_{D3d}$ = 2A$_{1g}$ + 2A$_{2g}$ + 4E$_g$ + 2A$_{1u}$ + 2A$_{2u}$ + 4E$_u$. Excluding the three acoustic modes (the doubly-degenerate E$_u$ and A$_{1u}$ modes), and noticing that each of the E$_g$ and E$_u$ modes are doubly degenerate, there are 21 modes in total. 

\begin{figure*}[ht]
 \includegraphics[width=\linewidth]{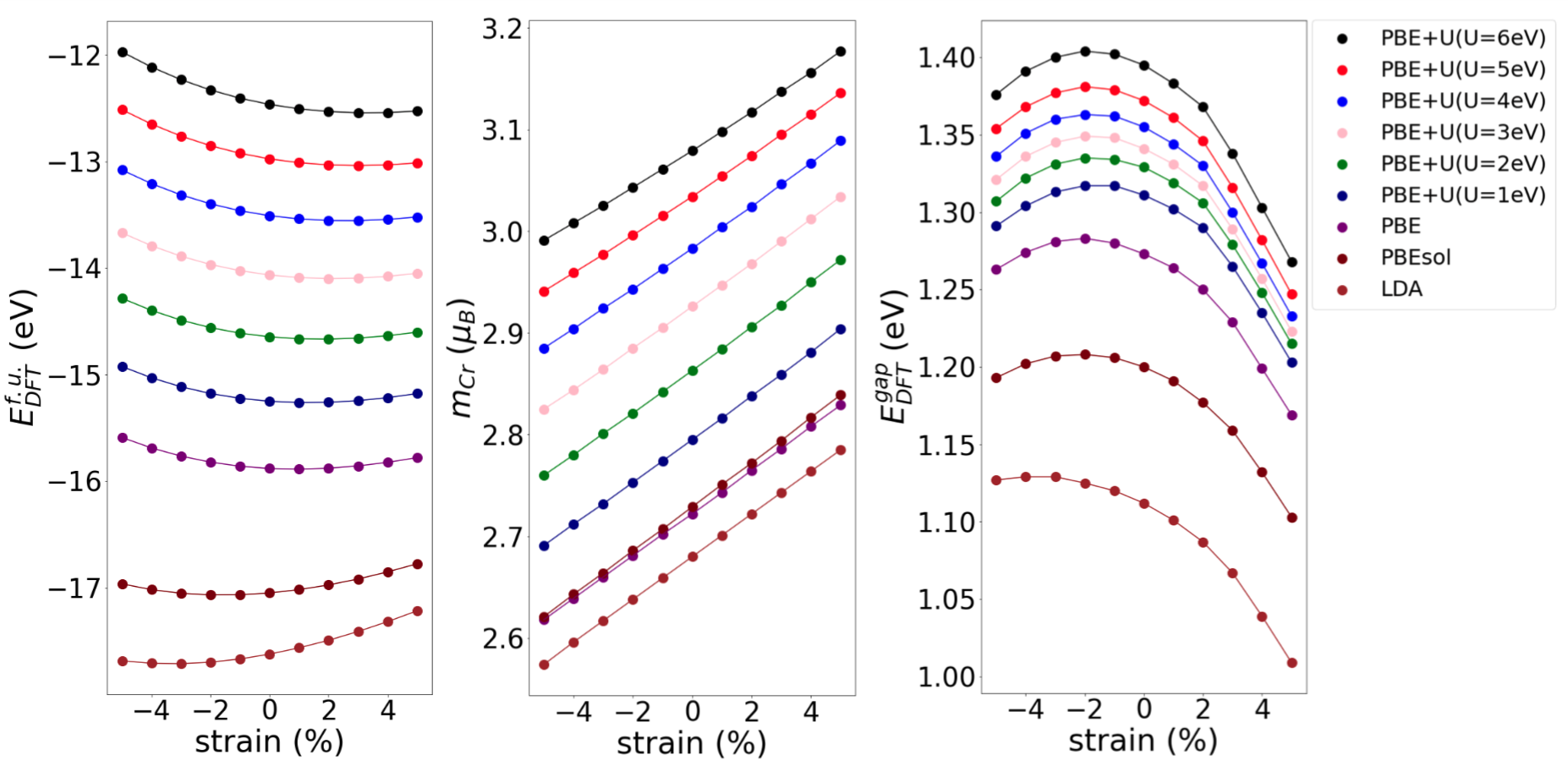}
 \caption{DFT predictions of the lattice constant, magnetic moment, and electronic band gap of bulk CrI$_3$ using different functionals. In the above, $E^{f.u.}$ designates the energy per formula unit. } 
 \label{fig:mock_DFT}
\end{figure*}

\section{Results and Discussions}
\label{Results}
  \vspace*{-0.10cm}

\subsection{Sensitivity of Monolayer Properties to Functional Choice \label{DFTU}} 



\begin{table}[b]
\centering
\caption{Tabulation of lattice parameters and magnetic moments predicted in this work compared to those from other studies. ($^{*}$ experimentally-determined in study of ML CrI$_3$ by Ref. \onlinecite{Li_ScienceBull}).}
\hspace*{-0cm}\begin{tabular}{||c c c c c ||}
\hline
 {\bf Ref.} & {\bf Method} & {\bf $a_0$ (Å)} & {\bf $m_{Cr} \; (\mu_B)$} & {\bf $m_{I} \; (\mu_B)$} \\ [0.5ex] 
 \hline\hline
\textit{This work} & LS-DMC & 6.87(3) & 3.61(9) & -0.14(5) \\ 
\hline 
\textit{This work} & LDA+$U$ & 6.695 & 3.497 & -0.099 \\
\hline 
\textit{Li}\cite{Li_ScienceBull} & GGA+$U$ & 6.84$^{*}$ & 3.28 & -- \\ 
\hline
\textit{Yang}\cite{Yang_JPCC} & GGA+$U$ & -- & 3.32  & --  \\
 \hline
 \textit{Wu}\cite{Wu_PCCP} & GGA+$U$ & 6.978 & 3.106  & --  \\
\hline
 \textit{Lado}\cite{Lado_PhysRevB} & DFT+$U$ & 6.686 & 3 & -- \\
\hline
 \textit{Zhang}\cite{Zhang_JMaterChemC} & PBE(HSE06) & 7.008 & 3.103 & -- \\ [1ex] 
 \hline
\end{tabular}
\label{tab:parameter_table1}
\end{table}

As a starting point for our subsequent DMC calculations, we began by modeling the monolayer's properties using standard density functionals, including the LDA, the GGA implementation of Perdew-Burke-Ernzenhof (PBE),\cite{perdew1996} PBEsol,\cite{perdew2008} and a series of PBE+$U$ functionals. Previous work has shown that the material properties can depend strongly on the functional employed and thus we initially sought a functional that could simultaneously predict the lattice constant, magnetic moment, and electronic band gap of the monolayer. Since relatively few \textit{ab initio} and experimental studies have been able to ascertain the properties of the monolayer, we fixed the internal geometry of the monolayer to the previously-determined bulk values. As illustrated in Figure \ref{fig:mock_DFT}, no single functional came close to simultaneously reproducing all three bulk properties: PBE+$U=1$ eV most closely approximated the bulk lattice constant, PBE+$U=6$ eV most closely approximated the bulk magnetic moment, and PBEsol most closely approximated the bulk band gap. Even more glaringly, all three properties continuously increased with the $U$ employed in our PBE+$U$ calculations, strongly suggesting that the $U$s employed could be continuously tuned to reproduce any individual quantity desired. This is illustrative of the sensitivity of CrI$_3$ properties to the variability in DFT functionals which motivates the use of DMC throughout the rest of this work.

To select a DFT functional that could serve as a meaningful reference, we thus turned to DMC simulations. Based on Figure \ref{fig:Monolayer_uValues}, LDA+$U$ wave functions with $U$s of ~2-3 eV minimize the DMC energy, so in all subsequent DFT analyses, we have employed LDA+$U$ wave functions with $U=2-3$ eV. This range of $U$ values is consistent with our linear-response estimate of $U_{LR}$=3.3 eV and also that employed in a wide range of materials and a recent DMC study of bulk CrI$_3$.\cite{Ichibha_PRM_2021}

With a $U=2$ eV value,  we used LDA+$U$ to relax the CrI$_3$ structure to determine its equilibrium geometry. As shown in Tables \ref{tab:parameter_table1} and \ref{tab:parameter_table2}, LDA+$U$ yields a lattice constant of 6.695~\r{A} and axial iodine-chromium-iodine bond angle ($\theta_{I-Cr-I}$) of 175.72°. The predicted lattice constant is within the range of lattice constants previously obtained using other density functionals, which range from 6.686~\r{A} from previous DFT+$U$ calculations\cite{Lado_PhysRevB} to 6.978~\r{A} from GGA+$U$ calculations,\cite{Wu_PCCP} a variation of roughly 5\%. In contrast, recent STM experiments point to an experimental monolayer lattice parameter of 6.84~Å, which is greater than our best LDA+$U$ estimate, a point to which we will return below. 

\subsection{DMC Predictions of the Monolayer Structure \label{LineSearch}} 

Given this variability in DFT-derived lattice constants, we employed DMC guided by a surrogate Hessian method to determine monolayer CrI$_3$'s equilibrium geometry with DMC-level accuracy. For monolayer CrI$_3$, the surrogate Hessian method yielded converged structural parameters to within the desired accuracy of 0.5\% of the lattice parameter within three iterations.

\begin{figure}[b]
 \includegraphics[width=3.45 in]{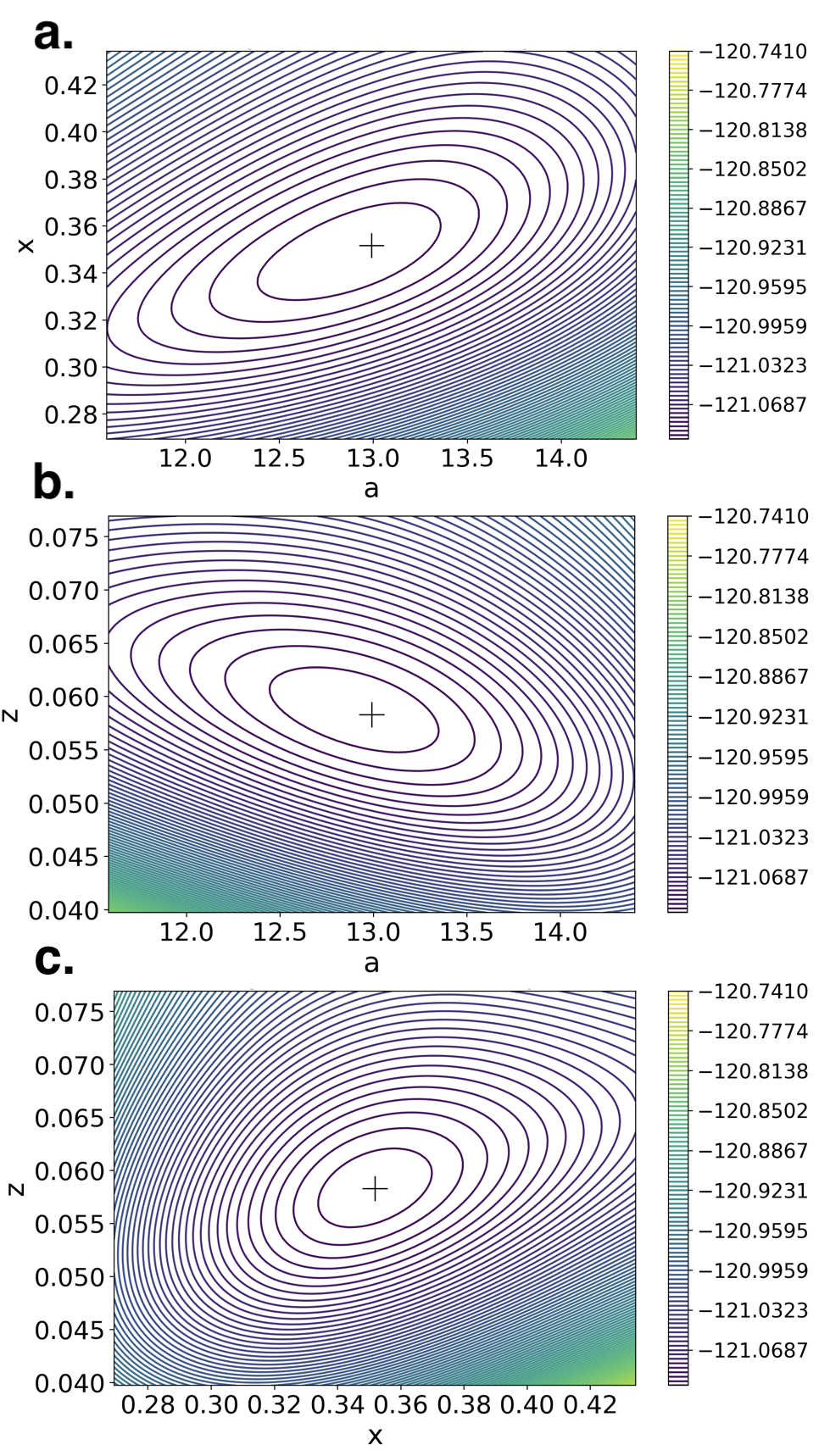}
 \caption{DMC potential energy surfaces obtained using the surrogate Hessian approach. The PES along the (a) ($a$,$x$) parameter pair plane, (b) ($a$,$z$) parameter pair plane, and (c) ($x$,$z$) parameter pair plane. Here, $a$, $x$, and $z$ denote Wyckoff parameters, of which $a \equiv a_0$ in Bohr and $x$ and $z$ combine to give bond angles and distances. The energy contours are separated by 4 mHa.} 
 \label{fig:contour_panel}
\end{figure}

As evidenced by the tightly-bunched contour lines in Figure \ref{fig:contour_panel}, the potential energy surface around the DMC structural minimum is steep and highly harmonic. In terms of the Wyckoff parameters $x$ and $a$ (note that $a \equiv a_0$), the DMC PES is more shallow along the direction in which $x$ and $a$ simultaneously increase, and steeper in the direction in which $x$ increases, but $a$ decreases. In contrast, the PES is steeper along the direction in which $z$ and $a$ simultaneously increase and more shallow along the direction in which $z$ decreases, but $a$ increases. Lastly, although the relationship between $z$ and $x$ is slightly more anharmonic than the previous two relationships, the PES roughly tends to be more shallow as both $z$ and $x$ increase, but steeper in the direction in which $z$ decreases, while $x$ increases. 

\begin{figure*}[t]
 \includegraphics[width=\linewidth]{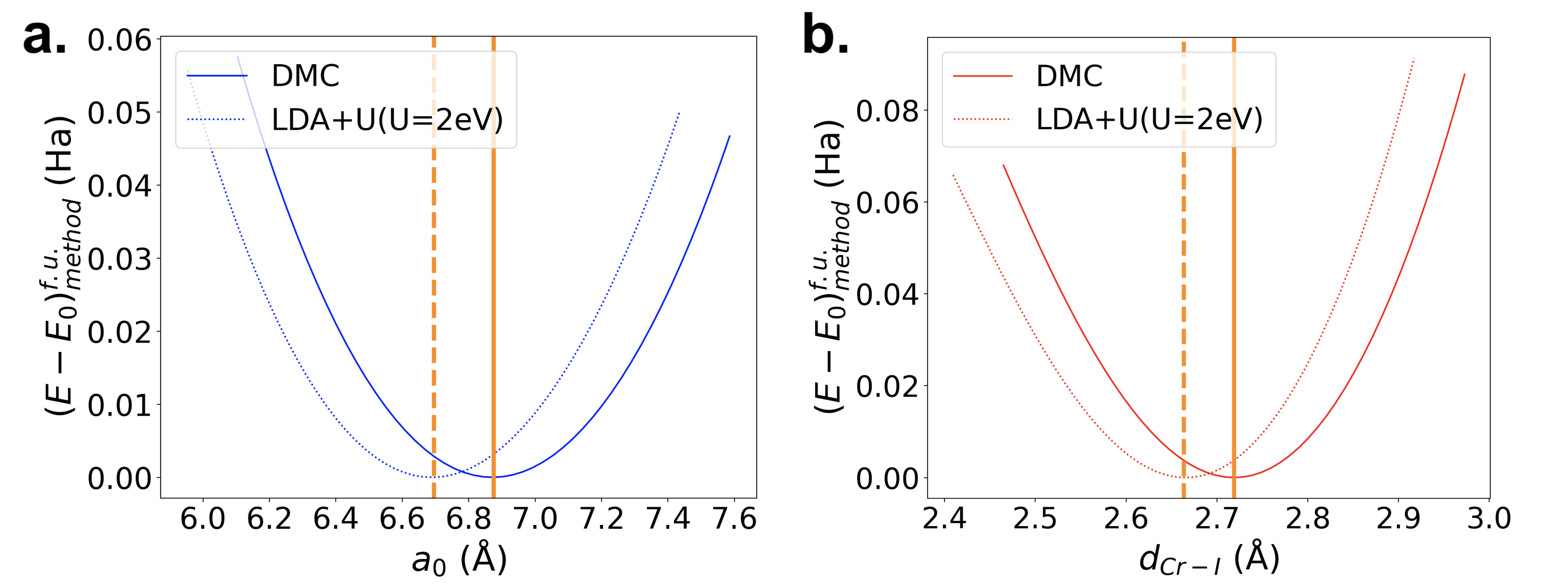}
 \caption{Relative energies for ML CrI$_3$ obtained using DMC and LDA+$U$ ($U=2$eV) vs. (a) lattice constant and (b) Cr-I bond distance. $E_{0}$ denotes the minimum energy per formula unit (f.u.) yielded by each method, with LDA+$U$ $E_0 = -185.3$ Ha and DMC $E_0 = -121.1$ Ha. The optimal lattice constant and Cr-I bond distance obtained by DMC are denoted by solid orange vertical lines, while those obtained by LDA+$U$ ($U=2$ eV) are denoted by dashed orange lines. }
 \label{fig:param_panel}
\end{figure*}

As bond angles and lengths consist of contributions from both the $x$ and $z$ Wyckoff parameters, a physical understanding of the DMC PES is enhanced by considering Fig.~\ref{fig:param_panel} of the main text and Figs.~\ref{fig:angle1_supp} and \ref{fig:angle2_supp} in the Supplementary Information. In these Figures, slices of the DMC PES near the true minima are taken so as to illustrate (i) how the CrI$_3$ energy changes with the lattice constant, $a_{0}$, Cr-I bond length, $d_{Cr-I}$, and CrI$_3$ bond angles, $\theta_1$ and $\theta_2$ (see Figure \ref{fig:mock_structure} for a visualization of how these quantities are defined) and (ii) that these structural quantities differ significantly between LDA$+U$ and DMC. These figures reveal that the energy changes most rapidly as the lattice constant is varied. The energy minima become increasingly more shallow along the $d_{Cr-I}$, $\theta_1$, and $\theta_2$ directions, signifying that it is more energetically costly to isotropically strain the lattice than to more locally vary bond lengths and angles. The relatively low barriers to changing the bond distances and angles are what moreover make it challenging to accurately resolve CrI$_3$'s structure, a feature that CrI$_3$ holds in common with many 2D materials, as we have illustrated using DMC in our previous works.\cite{Shin_PhysRevMats}

Ultimately, our line search converges upon an LS-DMC lattice parameter of $a_0=6.87$ \r{A}, as presented in Figure \ref{fig:param_panel} and Table \ref{tab:parameter_table1}. This lies within the range of previous DFT-derived estimates and is 2\% larger than the DFT+$U$ lattice constant discussed earlier. Nevertheless, what adds particular confidence to these results is that this independently-derived lattice parameter is just 0.4\% off from that recently obtained for a CrI$_3$ monolayer grown using molecular beam epitaxy and analyzed using scanning tunneling microscopy (see Li in Table \ref{tab:parameter_table1}).\cite{Li_ScienceBull} The fact that our surrogate Hessian approach was able to so closely reproduce an experimental value underscores both DMC and the surrogate Hessian approach's accuracy for CrI$_3$. In addition to obtaining the monolayer lattice parameter, these calculations also yielded estimates of the chromium-iodine bond distance ($d_{Cr-I}=2.723 Å$), as also presented in Figure \ref{fig:param_panel}, and the monolayer bond angles ($\theta_{1}=90.4°$ and  $\theta_{2}=175.4°$), as tabulated in Supplementary Table \ref{tab:parameter_table2} and Supplementary Figures \ref{fig:angle1_supp} and \ref{fig:angle2_supp}.

Although bond angles and distances have not been measured experimentally in the monolayer and are sparsely reported in DFT studies, our line search-predicted values all fall within 0.5\% of experimental values for bulk CrI$_3$. The other bond angles and distances obtained using different DFT functionals tabulated in Supplementary Table \ref{tab:parameter_table2} differ from the experimental bulk values by up to 3\%, further highlighting the robustness of our surrogate Hessian line search method for predicting structural parameters. 

\subsection{DMC Predictions of the Monolayer Magnetic Moment \label{DMC_moments}} 
\begin{figure*}[ht]
 \includegraphics[width=6.4 in]{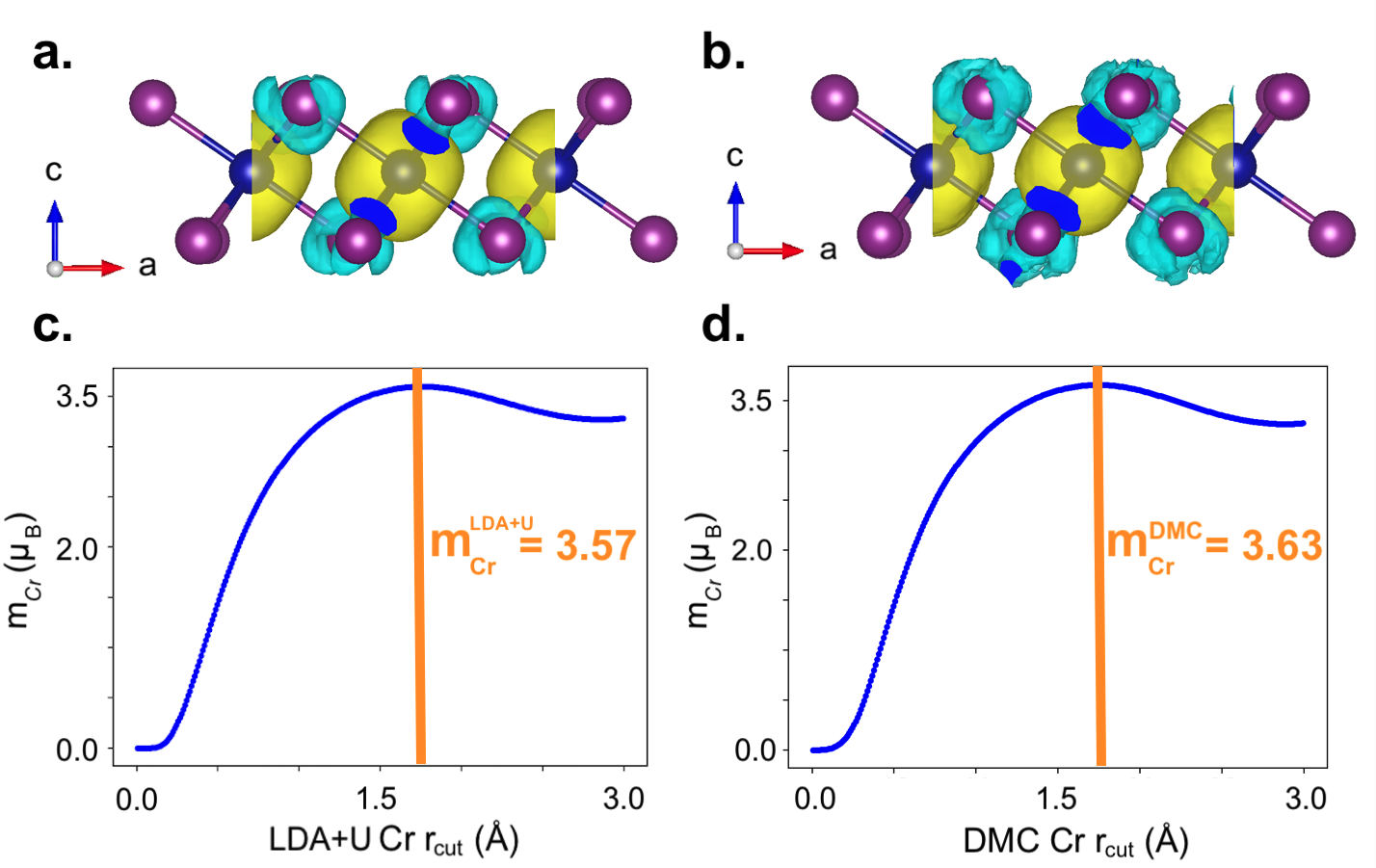}
 \caption{16-atom ML CrI$_3$ spin densities (isosurface = 0.00015 $(n_{up}-n_{down})/a_0^3$) for a monolayer cleaved from an experimental structure\cite{McGuire_ChemMater} obtained with (a) LDA+$U$ ($U=2$ eV) and (b) DMC. Magnetic moments of Cr as a function of the distance, $r_{cut}$, from the center of the Cr atom, obtained by interpolating over the (c) LDA+$U$ ($U=2$eV) and (d) DMC spin density grids. Areas with a net positive spin density are colored yellow while those with a net negative spin density are colored turquoise. The medium blue coloring is an artifact of the viewing angle. Magnetic moments are depicted by the orange lines, which are positioned at the maxima of the magnetic moment curves.} 
 \label{fig:momentFigure}
\end{figure*}

Using our DMC-optimized CrI$_3$ structure, we then proceeded to compute the magnetic moments on the Cr and I atoms in the hopes of fully and accurately resolving the magnetic structure of the monolayer for the first time.  By integrating over the DMC spin density, as described in Section \ref{MomentMeth}, we obtained a magnetic moment of 3.62 $\mu_{B}$ on each Cr and a moment of -0.15 $\mu_{B}$ on each I. Interestingly, this indicates the each Cr possesses a magnetic moment significantly larger than that of the the +3 charge that one would naively assume based upon chromium's typical oxidation state. Many previous DFT studies obtained an average magnetic moment of 3 $\mu_{B}$ across the entire CrI$_3$ unit cell, leading many researchers to conclude that all of the unit cell's magnetism could be attributed to that from the Cr site. 

While the sum of our magnetic moments for the Cr and four I atoms that constitute the unit cell also totals to 3 $\mu_{B}$, we instead find that the iodines additionally carry a significant magnetic moment, suggesting that CrI$_3$'s magnetism is more complex - and nonlocal - than previously assumed. As a check on these moments, we also employed LDA+$U$ and DMC to compute the moments of a monolayer cleaved from the experimental bulk structure.\cite{McGuire_ChemMater} The moments obtained using this structure were different by up to 0.06 $\mu_{B}$, supporting the assertion that DMC is capturing correlations missed by DFT (Figure \ref{fig:momentFigure}). Further, the DMC moments obtained using this structure were within 0.005 $\mu_{B}$ of those from our line search-optimized structure (see Table \ref{tab:parameter_table1}), a statistically insignificant discrepancy that strongly corroborates our findings. 

In contrast, integration over the spin densities of the LDA+$U$-optimized structure yielded magnetic moments of 3.497 $\mu_{B}$ on each Cr and -0.099 $\mu_{B}$ on each I, meaning that the correlation accounted for by DMC calculations leads to an increased spin anisotropy in the unit cell. That correlation leads to an increase in spin anisotropy is further substantiated by the fact that we see an increase in the magnitudes of the moments in the DMC-optimized structure to 3.655 $\mu_B$ on each Cr and -0.153 $\mu_B$ on each I after applying VMC to our DFT trial wave function, followed by a further tuning to their final values upon applying DMC to our VMC trial wave function (see Supplementary Figure \ref{fig:supp_moments}). 

\subsection{Spin-Phonon and Spin-Lattice Coupling \label{PhononResults}} 

The discrepancies in the magnetic moments we observe with varying monolayer geometries point to appreciable spin-phonon and spin-lattice couplings. Spin-phonon and spin-lattice couplings are associated with changes in the magnetic exchange interaction with changes in ionic-motion or strain, respectively, and together can give rise to a large magneto-elastic effect. A material with a large spin-phonon coupling can be leveraged for magneto-caloric device applications. 

Up to lowest order, the shift in the phonon frequency due to coupling to the lattice can be attributed to $\nu = \nu_0 + \lambda \langle S_i \cdot S_j \rangle$, where $\nu_0$ is the frequency in the paramagnetic case and $\lambda$ is the spin-phonon coupling constant with $i$ and $j$ as site indices. For CrI$_3$, if we assume a nominal +3 oxidation state for Cr, the expectation value of this spin-spin correlation function equates to $\sim 9/4$. Hence, for a given phonon frequency shift, we can estimate $\lambda$. The spin-spin correlation function is a constant for all modes in a given material. Hence, we simply report the frequency shifts in much of our discussion below. Nevertheless, we would like to note that for a given frequency shift, the effects of spin-anisotropy due to strong correlation, as revealed by our DMC calculations, should significantly increase the spin-spin correlation function, thereby effectively reducing the spin-phonon coupling.  

Experimentally, the largest spin-phonon coupling is observed in a 5$d$ perovskite oxide, corresponding to a frequency shift of 40 cm$^{-1}$.\cite{son_unconventional_2019}  In comparison, a frequency shift of $\sim$2.7 cm$^{-1}$ is observed for the E$_g$ mode of the chromium-based 2D material Cr$_2$Ge$_2$Te$_6$ (CGT) around its T$_c$. This corresponds to a $\lambda$ of $\sim1.2$ cm$^{-1}$. Spin-phonon couplings in the range of $0.27-0.4$ cm$^{-1}$ have subsequently been observed in other transition-metal trihalides, again arising from the E$_g$ modes.\cite{kozlenko2021spin} 

In our phonon calculations, we initially fixed CrI$_3$'s lattice parameter to its bulk value. 
Considering AFM ordering as one specific realization of the paramagnetic phase, we find that the largest change in phonon frequency between the AFM and FM phases is for the E$_g$ symmetric mode, and is $\sim$4 cm$^{-1}$ (see Table \ref{tab:spin-phonon}, modes 8 and 9). This is consistent with CrI$_3$ having a sizable spin-phonon coupling at finite temperatures, comparable to that of CGT and other 2D, layered magnetic materials.  

We next performed full geometry relaxations (using LDA+$U=3$ eV in VASP) for the different magnetic configurations. We find the optimal lattice parameters for the FM, AFM, and NM phases of the CrI$_3$ monolayer to be 6.677 \AA , 6.656 \AA, and 6.617 \AA, respectively (see Table \ref{tab:spin-phonon}). A large, $\sim$1 \% difference in lattice-parameters due to magnetic ordering is indicative of a non-trivial spin-lattice coupling and is consistent with unstable, in-plane acoustic modes when ML-CrI$_3$ is forced to be in a non-magnetic state (see Fig. \ref{fig:bands_DFT}). Unstable acoustic modes have been observed in certain Heusler compounds.\cite{zayak2005anomalous} A strong spin-lattice coupling, particularly involving soft in-plane acoustic modes, should lead to structural transitions across the Curie temperature. Indeed, bulk CrI$_3$ has been experimentally shown to transition from a rhombohedral to a monoclinic structure across the paramagnetic transition, \cite{McGuire_ChemMater} consistent with these findings. 

We also find that the phonon frequencies shift to larger values across all of the different magnetic orderings when the lattice parameter is allowed to relax, as shown in Table~\ref{tab:spin-phonon}. Specifically, the phonon frequency change between the FM and AFM phases increases to $\sim$10 $cm^{-1}$ for the E$_g$ mode (modes 8 and 9). This particular mode resembles shearing of the iodine planes, and is the one with the largest spin-phonon coupling in CGT as well as in other Cr trihalides. Assuming a spin-eigenvalue ${\bf S}$ of 3.47$\hbar$/2 based upon our DMC calculations of the magnetic moment, this corresponds to a $\lambda$ of $\sim3.32$ cm$^{-1}$. In comparison, the theoretically predicted spin-phonon coupling for CGT using a rigorous perturbation theory approach was 3.19 cm$^{-1}$.\cite{PhysRevB.100.224427}   

\begin{table}[t]
\centering
\caption{Frequencies of the phonon modes for the FM, AFM, and NM phases of the CrI$_3$ monolayer using the bulk and relaxed (r) lattice parameters predicted with the LDA+$U=3$ eV functional in VASP. Frequencies are specified in units of cm$^{-1}$.}
\begin{tabular}{l l l l l l l l} 
 \hline\hline
       Mode & Symm.  & FM & FM$_r$ & AFM & AFM$_r$ & NM & NM$_r$ \\
 \hline
    1,2 & E$_g$     & 47.6 & 50.8 & 47.3 & 49.6 & 39.9, 40.4 & 19.1, 41.6 \\  
      3 & A$_{2u}$ & 49.7 & 57.1 & 50.4 & 58.8 & 52.2 & 47.7 \\
      4 & A$_{1g}$ & 69.6 & 76.9 & 69.8 & 77.8 & 53.2 & 53.6  \\
    5,6 & E$_u$     & 78.2 & 81.2 & 79.2 & 82.9 & 74.5, 87.6 & 60.8, 71.8  \\
      7 & A$_{2g}$ & 84.5 & 88.7 & 86.3 & 90.7 & 91.4 & 82.6  \\
    8,9 & E$_g$     & 102  & 103  & 97.9 & 92.5 & 94.7, 104 & 92.4, 94.7  \\
 10, 11 & E$_g$     & 107  & 109  & 108  & 110  & 107, 110 & 103, 112  \\
 12, 13 & E$_u$     & 109  & 116  & 111  & 118  & 115, 120 & 115, 118 \\
     14 & A$_{1g}$ & 129  & 131  & 126  & 128  & 140 & 173  \\
     15 & A$_{2u}$ & 130  & 135  & 131  & 136  & 162 & 194  \\
     16 & A$_{2g}$ & 212  & 219  & 218  & 222  & 228 & 205  \\
 17, 18 & E$_u$     & 221  & 229  & 227  & 235  & 231, 232 & 206, 222  \\
 19, 20 & E$_g$     & 238  & 245  & 236  & 235  & 241, 251 & 236, 254 \\
     21 & A$_{1u}$ & 258  & 267  & 251  & 261  & 313 & 298  \\
 \hline
\end{tabular}
\label{tab:spin-phonon}
\end{table}

These results demonstrate that even in the monolayer limit, CrI$_3$ possesses strong spin-phonon and spin-lattice couplings. While recent studies have demonstrated a magneto-optic effect in few-layer CrI$_3$, the large coupling of the magnetism to the Raman active phonon frequencies and lattice-strain that we observe in this study suggest that one could potentially use magnetic fields to modulate inelastically scattered light.  Indeed, a very recent experiment demonstrates that an out-of-plane magnetic field change from -2.5~T to 2.5~T leads to a rotation in the plane of the polarization of inelastically scattered light from -20$^\circ$ to +60$^\circ$.\cite{liu2020observation} 

\section{Conclusions}
\label{Conc}

In summary, we have used a potent combination of first principles DFT and DMC calculations to produce some of the most accurate estimates of the electronic, magnetic, and structural properties of monolayer CrI$_3$ to date. Using a surrogate Hessian line search optimization technique combined with DMC, we were able to independently resolve the lattice and other structural parameters of CrI$_3$ to within a fraction of a percent of recently-published STM measurements, an accomplishment given the up to 10\% variability in previous DFT-derived estimates of the lattice parameter depending upon the functional employed. Based upon the DMC-quality structure we obtained, we were then able to acquire a high-resolution monolayer spin density that showed each Cr atom to possess a magnetic moment of 3.62 $\mu_B$ and each I atom to have a moment of -0.145 $\mu_B$, substantially larger moments than previously reported that suggest that CrI$_3$ manifests substantial ligand magnetism. Given the Cr-I-Cr angle of 90$^\circ$, this would indicate a superexchange stabilization of the ferromagnetic ordering. In conjunction with the expected large spin-orbit coupling on the I atom,\cite{Lado_2017} this could also give rise to a large ligand superexchange-dominated magnetic-anisotropy,\cite{PhysRevLett.122.207201, PhysRevLett.124.017201} explaining recent observations of spin-waves and magnons in thin-film CrI$_3$.\cite{cenker_direct_2021} We moreover demonstrated that CrI$_3$ possesses remarkably strong spin-phonon coupling, with a predicted value as large as 3.32 cm$^{-1}$. This work thus demonstrates CrI$_3$'s promise for magnon-based spintronic applications,\cite{chumak_magnon_2015} potentially with optical controls,\cite{liu2020observation} while also demonstrating the capability of DMC to accurately model the structural and magnetic properties of 2D materials.

\section{Acknowledgments}
\label{Ack}

The authors thank Paul Kent, Kemp Plumb, Anand Bhattacharya, Ho Nyung Lee, Fernando Reboredo, Nikhil Sivadas and the QMCPACK team for thoughtful conversations. The work by G.H., J.T., R.N., J.K., M.C.B., O.H., P.G., and B.R., and D.S.'s modeling and analysis efforts, and the scientific applications of QMCPACK were supported by the U.S. Department of Energy, Office of Science, Basic Energy Sciences, Materials Sciences and Engineering Division, as part of the Computational Materials Sciences Program and the Center for Predictive Simulation of Functional Materials. D.S.'s work preparing this manuscript was supported by the NASA Rhode Island Space Grant Consortium. This research was conducted using computational resources and services at the Center for Computation and Visualization, Brown University and the National Energy Research Scientific Computing Center (NERSC), a U.S. Department of Energy Office of Science User Facility operated under
Contract No. DE-AC02-05CH11231. \\
    
$^\dag$ D. S. and G. H. made equal contributions as first author to the manuscript. 
$^{\dag\dag}$ Corresponding authors   B. R. (brenda\underline{ }rubenstein@brown.edu) \&  P. G. (ganeshp@ornl.gov). 

\section{Data availability}

The data that supports the findings of this study are available within the article and its supplementary material.

\bibliography{ref}

\makeatletter\@input{xx.tex}\makeatother
\end{document}


\title{Supplementary Materials for ``A Combined First Principles Study of the Structural, Magnetic, and Phonon Properties of Monolayer CrI$_{3}$"}
\author{Daniel Staros,$^{1, \dag}$ Guoxiang Hu,$^{2, 3, \dag}$ Juha Tiihonen,$^{6}$ Ravindra Nanguneri,$^{1}$ Jaron Krogel,$^{6}$ M. Chandler Bennett,$^{6}$ Olle Heinonen,$^{4,5}$ Panchapakesan Ganesh,$^{3, \dag\dag}$ and Brenda Rubenstein$^{1,\dag\dag}$}

\address{$^{1}$Department of Chemistry, Brown University, Providence, RI 02912, USA}
\address{$^{2}$Department of Chemistry and Biochemistry, Queens College, City University of New York, Flushing, NY 11367, USA}
\address{$^{3}$Center for Nanophase Materials Sciences Division, Oak Ridge National Laboratory, Oak Ridge, TN 37831, USA}
\address{$^{4}$Materials Science Division, Argonne National Laboratory, Argonne, IL 60439, USA}
\address{$^{5}$Northwestern-Argonne Institute for Science and Engineering, Northwestern University, Evanston, IL 60208, USA}
\address{$^{6}$Material Science and Technology Division, Oak Ridge National Laboratory, Oak Ridge, TN 37831, USA}

\date{\today}
\maketitle 

\section{Monolayer Structural Parameters} \label{supp-introduction}
  \vspace*{-0.20cm}

\renewcommand{\thefigure}{SI}
\begin{figure}[b]
 \includegraphics[width=\linewidth]{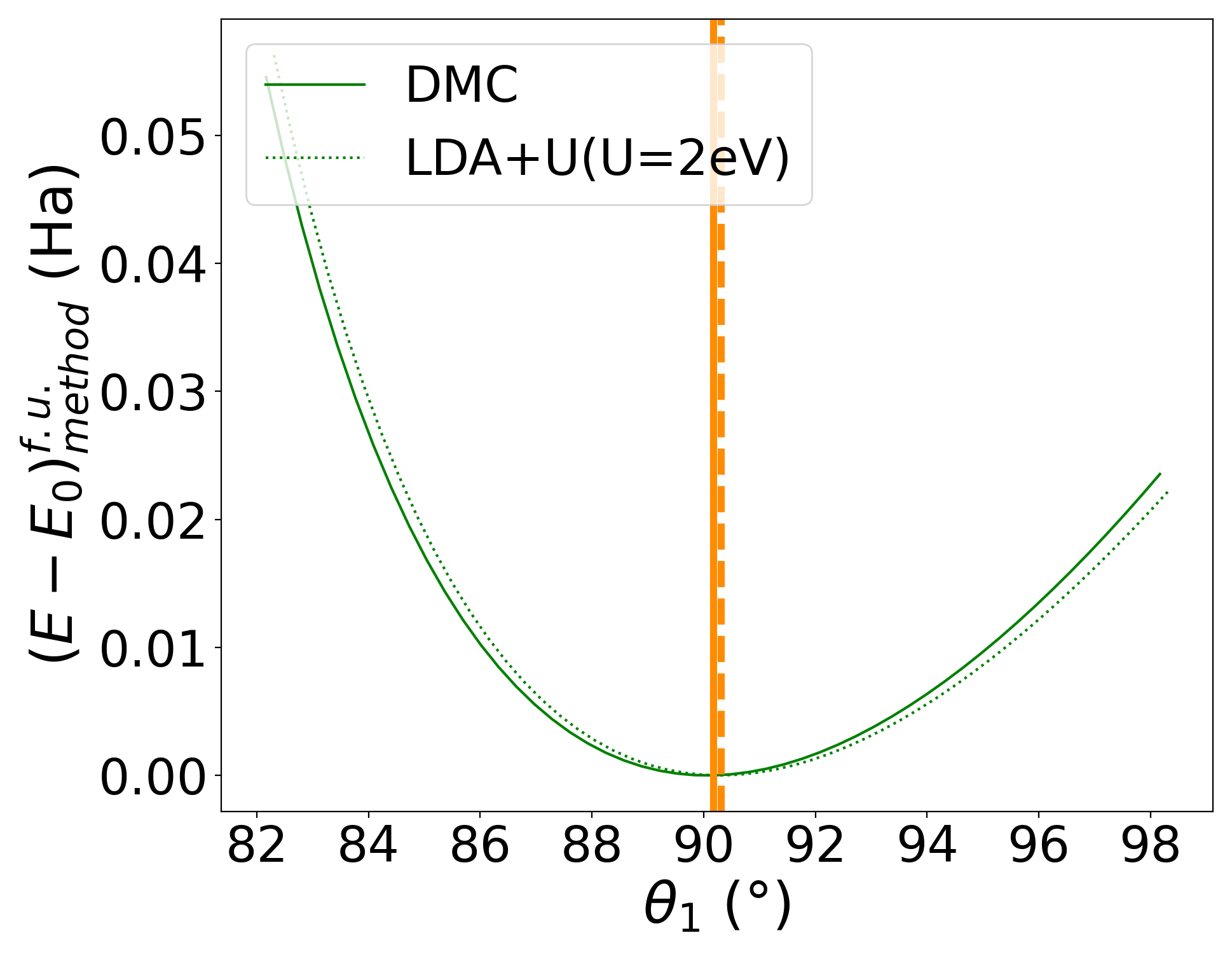}
 \caption{Relative energies for ML CrI$_3$ obtained using DMC and LDA+$U$ ($U$=2~eV) vs. $\theta_1$ bond angle. $E_{0}$ denotes the minimum energy per formula unit (f.u.) yielded by each method, with LDA+$U$ having an $E_0 = -185.3$~Ha and DMC having an $E_0 = -121.1$~Ha. The optimal physical parameters obtained by DMC are denoted by solid orange vertical lines while those obtained by LDA+$U$ ($U=2$~eV) are denoted by dashed orange lines.}
 \label{fig:angle1_supp}
\end{figure}
In this section, we provide additional results regarding the structural parameters of the optimal structures obtained via the surrogate Hessian line search method and LDA+$U$. Figures \ref{fig:angle1_supp} and \ref{fig:angle2_supp} respectively show slices of the LDA+$U$ and DMC potential energy surface along the directions of the equatorial ($\theta_1$) and axial ($\theta_2$) I-Cr-I bond angles (illustrated in Figure 1 of the main text). The angle $\theta_1$ exhibits minima corresponding to the LDA+$U$ and DMC-optimized structures at 90.311° and 90.169°, respectively, while the angle $\theta_2$ exhibits minima corresponding to the LDA+$U$ and DMC-optimized structures at 176.057° and 175.236°, respectively. Both $\theta_1$ and $\theta_2$ are smaller for the LDA+$U$ case than for the DMC case.

\renewcommand{\thefigure}{SII}
\begin{figure}[b]
 \includegraphics[width=\linewidth]{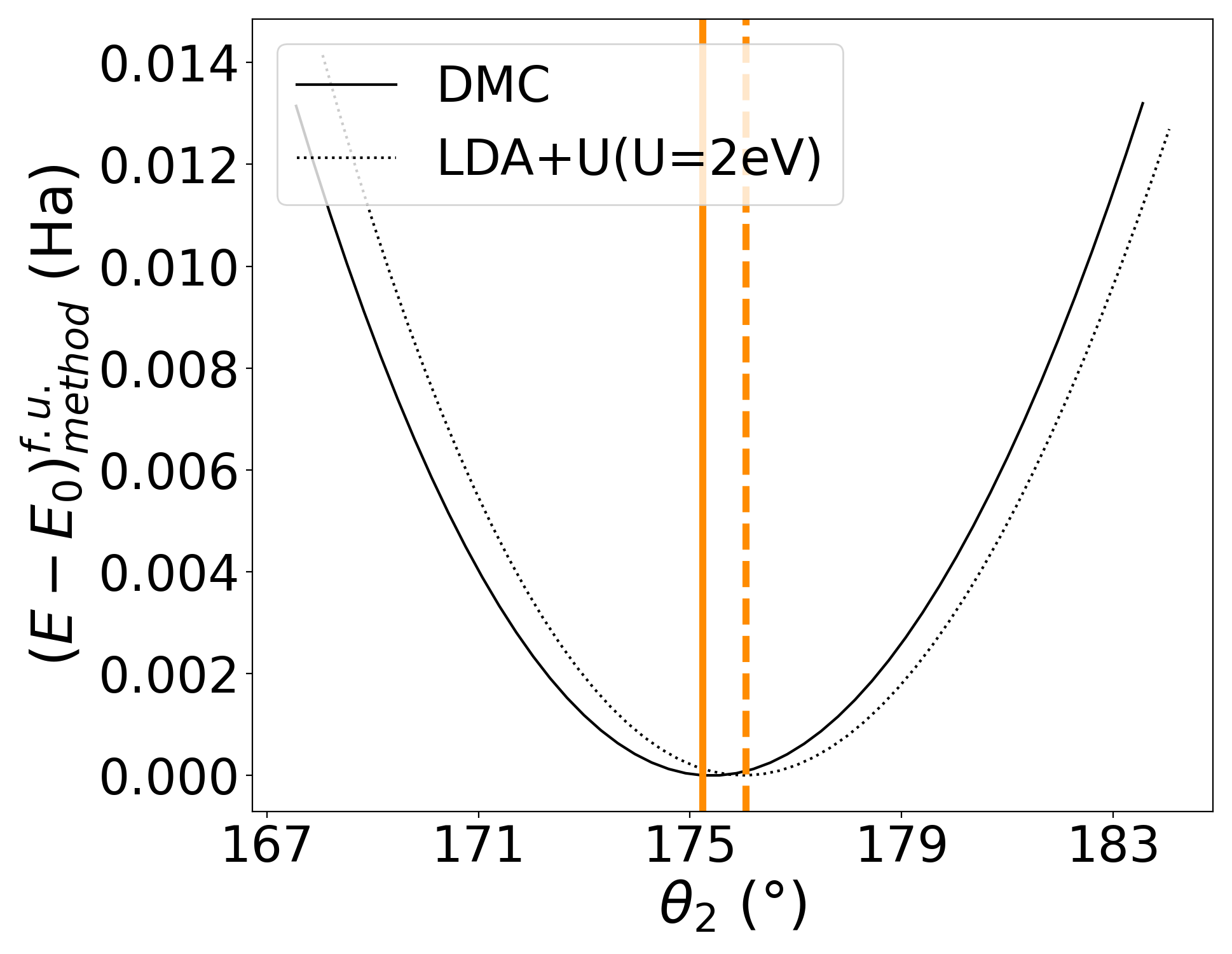}
 \caption{Relative energies for ML CrI$_3$ obtained using DMC and LDA+$U$ ($U$=2 eV) vs. $\theta_2$ bond angle. $E_{0}$ denotes the minimum energy per formula unit (f.u.) yielded by each method, with LDA+$U$ having an $E_0 = -185.3$ Ha and DMC having an $E_0 = -121.1$ Ha. The optimal physical parameters obtained by DMC are denoted by solid orange vertical lines, while those obtained by LDA+$U$ ($U$=2 eV) are denoted by dashed orange lines.}
 \label{fig:angle2_supp}
\end{figure}

We also tabulate methodological details (Method, Hubbard $U$), structural parameters (lattice constant $a_0$, bond lengths $d_{Cr-I}$, and bond angles $\theta_1$ and $\theta_2$) and magnetic moments ($m_{Cr}$ and $m_{I}$) obtained by previous studies alongside the findings of this work in \ref{tab:parameter_table2}. As illustrated in this table, DFT results for the lattice parameter vary by upwards of 5\%. CrI$_3$ bond lengths and angles can vary by a few percent, but comparatively few works have reported these values. Published magnetic moments on the Cr atom vary by an even greater factor, up to 20\%. 

\begin{table*}[ht]
\renewcommand\thetable{AI}
\caption{Optimal structural parameters and predicted atomic magnetic moments by reference for ML CrI$_3$ ($^{*}$ = experimentally determined for ML CrI$_3$, $^{**}$ = experimentally determined for bulk CrI$_3$, $^{***}$ = SOC included). Note that our LDA+$U$ relaxation utilized ultrasoft non-norm conserving pseudopotentials, but the SCF to get moments utilized the harder TM pseudopotentials.}
\hspace*{-0cm}\begin{tabular}{||c c c c c c c c c ||}
\hline
 Ref. & Method & $U$ (eV)& $a_0$ (Å) & $d_{Cr-I}$ (Å)  & $\theta_{1} \degree$ & $\theta_{2} \degree$ & $m_{Cr} \; (\mu_B)$ & $m_{I} \; (\mu_B)$ \\ [0.5ex] 
 \hline\hline
\textit{This work} & LS-DMC & 2.0 & 6.87(3) & 2.722(8)  & 175.4(3) & 90.4(2) & 3.619 (0.003) & -0.145 (0.002) \\ 
\hline 
\textit{This work} & LDA+$U$ & 2.0 & 6.695 & 2.659 & 175.72 & 90.26 & 3.497 & -0.099 \\
\hline 
\textit{Li}\cite{Li_ScienceBull} & GGA+$U$ & 3.9 & 6.84$^*$ & --  & -- & -- & 3.28 & -- \\ 
\hline
\textit{McGuire (bulk)}\cite{McGuire_ChemMater} & -- & -- & 6.867$^{**}$ & 2.726$^{**}$  & 175.69$^{**}$ & 90$^{**}$ & -- & -- \\
\hline 
\textit{Katsnelson}\cite{Katsnelson_PRB} & DFT+$U$ & 3 & 6.842 & 2.705 & -- & -- & --  & --  \\
\hline
\textit{Katsnelson}\cite{Katsnelson_PRB} & sDFT+$U$ & 3 & 6.856 & 2.710 & -- & -- & --  & --  \\
\hline
\textit{Katsnelson}\cite{Katsnelson_PRB} & sDFT & -- & 6.817 & 2.690 & -- & -- & -- & -- \\
 \hline
\textit{Kashin}\cite{Kashin_2DMats} & GGA+$U$ & 3 & 4.03 & --   & -- & -- & 3.119  & --  \\
\hline
\textit{Yang}\cite{Yang_JPCC} & GGA+$U$ & 2 & -- & --   & -- & -- & 3.32  & --  \\
 \hline
 \textit{Wu}\cite{Wu_PCCP} & GGA+$U$ & 2.7 & 6.978 & --   & -- & -- & 3.106  & --  \\
\hline
\textit{Behera}\cite{Behera_AppPhysLett} & GGA+$U$ & 3 & 6.963 & 2.76  & -- & -- & -- & -- \\
\hline
 \textit{Webster(1)}\cite{Webster_PhysChemChemPhys} & PBE & -- & 7.008 & 2.740  & 173.3 & -- & -- & -- \\ 
 \hline
 \textit{Webster(2)}\cite{Webster_PhysRevB} & LDA$^{***}$ & -- & 6.689 & 2.655  & 175.7 & 90.2 & -- & -- \\
\hline
 \textit{Lado}\cite{Lado_PhysRevB} & DFT+$U^{***}$ & 2.7 & -- & --  & -- & -- & 3 & -- \\
\hline
 \textit{Zhang}\cite{Zhang_JMaterChemC} & PBE & -- & 7.008 & --  & -- & -- & 3.103 & -- \\
 \hline
 \textit{Zhang}\cite{Zhang_JMaterChemC} & optB88-vdW & -- & 6.905 & --  & -- & -- & -- & -- \\
 \hline
 \textit{Zhang}\cite{Zhang_JMaterChemC} & HSE06 & -- & -- & --  & -- & -- & 3.305 & -- \\
 [1ex] 
 \hline

\end{tabular}
\vspace{4mm}
\label{tab:parameter_table2}
\end{table*}

\renewcommand{\thefigure}{SIII}
\begin{figure*}[ht]
 \includegraphics[width=\linewidth]{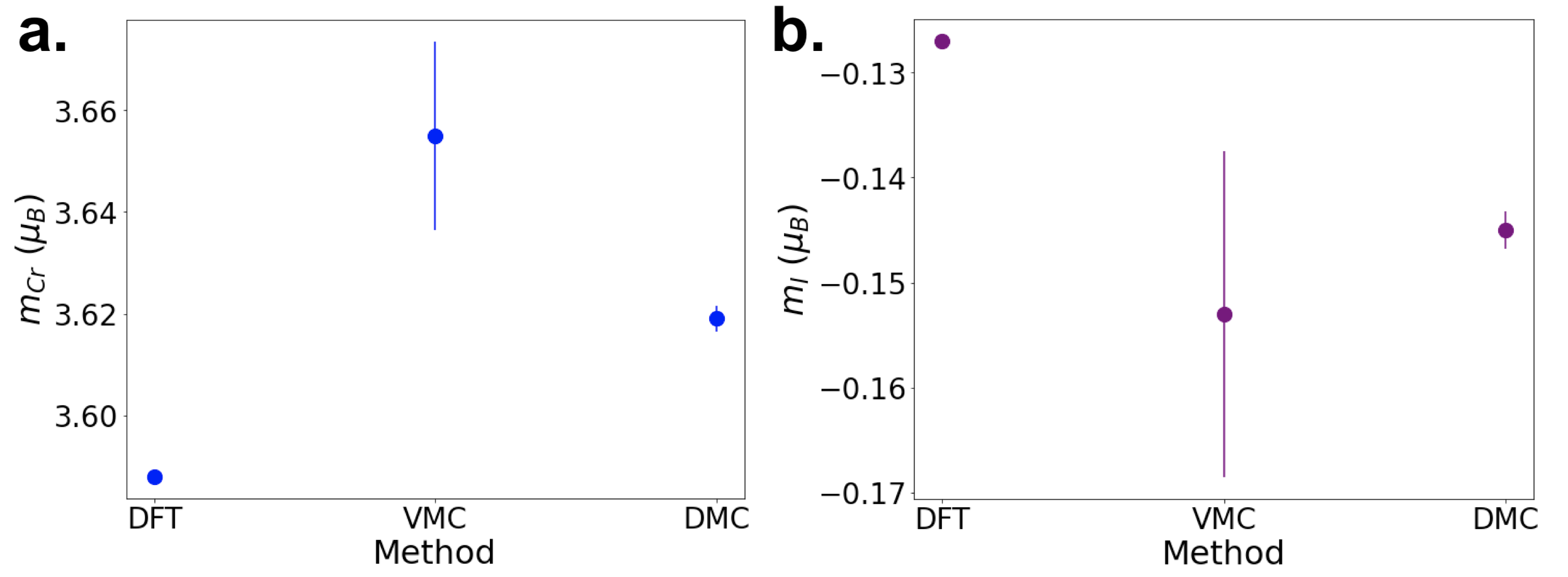}
 \caption{The change in predicted atomic magnetic moments on (a) chromium and (b) iodine in the DMC-optimized structure as obtained from DFT, VMC, and DMC spin densities. Note that the magnitudes of the moments on both atoms obtained using VMC seem to increase more than expected based upon the final DMC results. The DMC values are still larger in magnitude than the DFT values.}
 \label{fig:supp_moments}
\end{figure*}

The work of Li \textit{et al.} is one of the first experimental works to provide high-accuracy structural parameters for ML CrI$_3$ and the work of McGuire \textit{et al.} is the first to provide structural parameters for the bulk solid. Despite the variability in the values reported in this table, it should be noted that our LS-DMC results agree with the two experimental values reported to within a fraction of a percent. This is remarkable given that our simulations were conducted without knowledge of these experimental results. The structural parameters presented here thus represent one of the most comprehensive - and accurate - sets presented to date for this material.
 
\section{Prediction of Magnetic Moments}
\label{supp-moments}

\renewcommand{\thefigure}{SIV}
\begin{figure*}[t]
 \includegraphics[width=\linewidth]{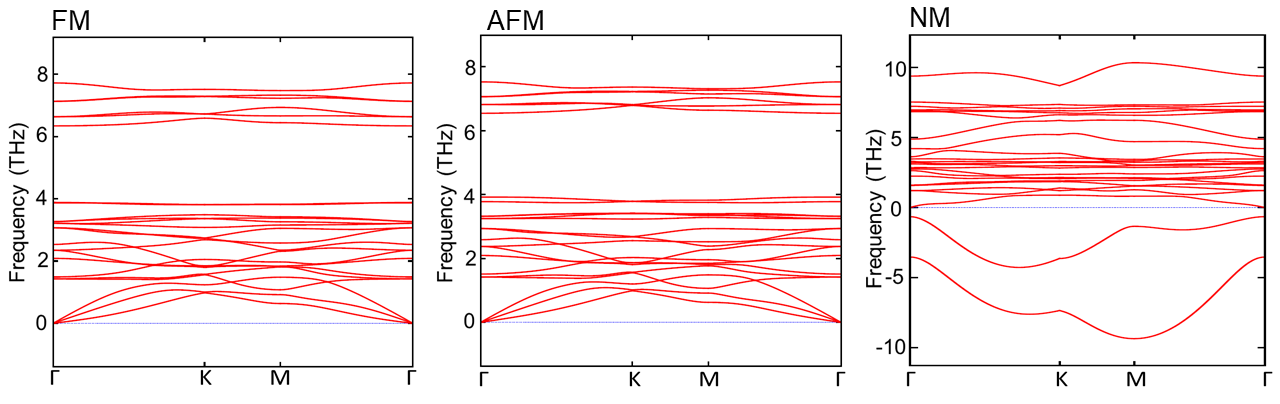}
 \caption{Phonon bands obtained for ferromagnetic, antiferromagnetic, and non-magnetic ML CrI$_3$ from linear-response LDA+U=3eV.} 
 \label{fig:bands_DFT}
\end{figure*}

\renewcommand{\thefigure}{SV}
\begin{figure}[t]
 \includegraphics[width=3.2 in]{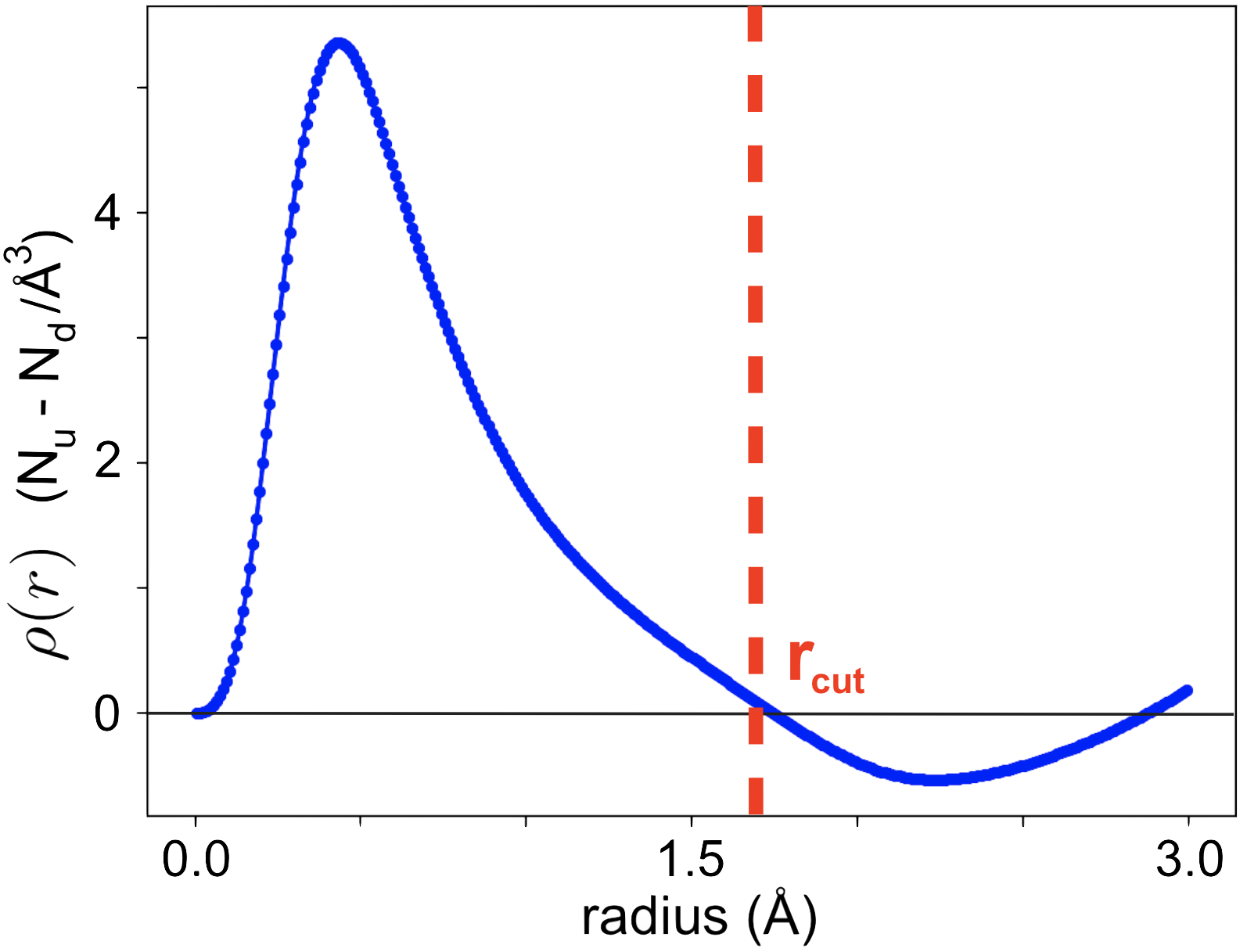}
 \caption{Example plot of a DMC radial spin density as a function of radius from the Cr atom, with the location of zero-recrossing denoted by a red dashed line.}
 \label{fig:rcut_figure}
\end{figure}

In Figure \ref{fig:supp_moments}, we illustrate how the predicted magnetic moments vary after each of the DFT, VMC, and DMC steps of the DMC workflow. Note that all of these moments are computed using the DMC-optimized structure. The error bars on the VMC and DMC magnetic moments are the result of resampling, which consisted of generating several hundred spin density files according to the distribution of spin density errors originally output by the qdens capability of QMCPACK, followed by a statistical analysis of the corresponding atomic moments and standard deviations. The increase in the magnitude of the atomic moments moving from DFT to VMC demonstrates the additional electron correlation resolved by VMC. A slight decrease is observed moving from VMC to DMC. The large VMC error bars are a consequence of the smaller number of samples used for VMC than those used for DMC. 

\section{Spin-Phonon and Spin-Lattice Coupling} \label{supp-phonon}

The phonon spectra of ferromagnetic (FM), antiferromagnetic (AFM), and nonmagnetic (NM) monolayer CrI$_3$ obtained via linear response are presented in Figure \ref{fig:bands_DFT}. Significant differences can be observed between the three phases. The nonmagnetic spectrum exhibits modes between 0 and 6 THz which are not present in the FM or AFM spectra; this feature of the NM spectrum is indicative of unstable in-plane acoustic modes, as discussed in the main text.

\section{Magnetic moment cutoff radius} \label{supp-phonon}
The cutoff radius we employed for all moment calculations corresponded to the zero recrossing point of the spin density. This is illustrated in Figure \ref{fig:rcut_figure} for a DMC spin density. We chose to use this metric due to the fact that it has a physical basis as the point at which the spin density is no longer positive and eliminates the use of more arbitrary cutoffs. This definition is also flexible in that it can account for slight changes in spin density curves that arise in different structures.

\bibliography{ref}